\documentstyle[seceq,epsf,color]{ptptex}

\title{Kaonic Nuclear Clusters \\
--- a New Paradigm of Particle and Nuclear Physics ---}

\author{Yoshinori {\sc Akaishi}$^{*}$ and Toshimitsu {\sc Yamazaki}$^{**}$}

\inst{$^{*}$High-Energy Accelerator Research Organization (KEK), Tsukuba 305-0801, Japan \\
$^{**}$Department of Physics, University of Tokyo, Tokyo 113-0033, Japan }

\recdate{\today}

\abst{$\Lambda^*$=$\Lambda(1405)$ plays an essential role in forming kaonic nuclear clusters (KNC). The simplest KNC, $K^-pp$, has the structure $\Lambda^*p$=$(K^-p)^{I=0}p$, where a real kaon migrates between two nucleons, mediating super-strong $\Lambda^*p$ attraction. Production data of $K^-pp$ have been accumulated by DISTO, J-PARC E27 and J-PARC E15 experiments. In the case of $K^-K^-pp$, the attractive covalent bond of $\Lambda^*\Lambda^*$ is doubly enhanced compared to the $\Lambda^*p$ one. As a consequence, a $m$-th $\Lambda^*$ multiplet, $(\Lambda^*)_m$, with $m(m-1)/2$ bonds becomes more stable than its corresponding neutron aggregate, $(n)_m$, with $m=8 \sim 12$. This may suggest the possible existence of stable $\Lambda^*$ matter. The production of $K^-K^-pp$ by high-energy $pp$ or heavy-ion collisions is awaited as a doorway to so-far unknown $\Lambda^*$ matter.}

\begin{document}
\maketitle

\section{Introduction} 

In 2002, the present authors predicted the possible existence of deeply-bound anti-kaonic nuclear clusters (KNC), $K^-pp$ and $K^-ppn$, based on a phenomenological $\bar KN$ interaction derived from the $\Lambda^*=\Lambda(1405)=(K^-p)^{I=0}$ ansatz \cite{Akaishi02,Yamazaki02}. Hereafter, anti-kaon ($\bar K$) is referred to as "kaon" for simplicity. $\Lambda^*=\Lambda(1405)$ plays an essential role in forming KNC's. A microscopic variational calculation found that $K^-pp$ has a structure of $\Lambda^*p=(K^-p)^{I=0}p$ \cite{Yamazaki07,Akaishi10}, and the migration of a real kaon between nucleons provides an attractive super-strong nuclear force \cite{Yamazaki07b} between $\Lambda^*$ and $p$ by molecular bonding, which can be called the Heitler-London\cite{Heitler27}-Heisenberg\cite{Heisenberg32} mechanism. 
\vspace{-0cm}

\begin{figure}[hbtp]
\begin{center}
\hspace{0.5cm}
\vspace{-0.5cm}
\epsfxsize=13cm
\epsfbox{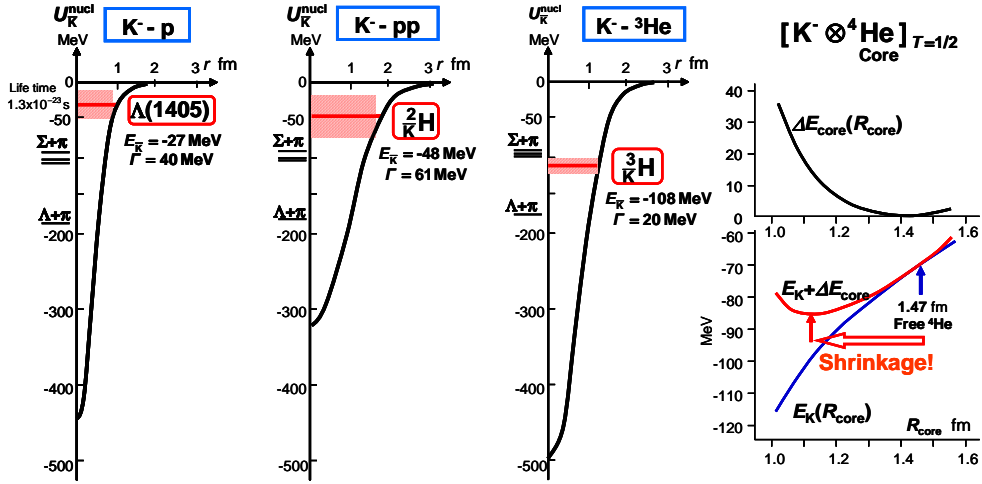}
\vspace{2cm}
\end{center}
\caption{Predicted kaonic nuclear cluster (KNC) states shown together with $\bar KN$ and $\bar K$-nucleus potentials. Strong shrinkage of KNC is demonstrated \cite{Akaishi02}.
\vspace{1cm}}
\label{fig1}
\end{figure}

A remarkable feature of KNC is a strong shrinkage of the system, \cite{Akaishi02} as demonstrated \cite{Dote02} in Fig. \ref{fig1}. The density goes up by several times higher than the normal nuclear density, $\rho_0=0.17~ {\rm fm}^{-3}$, due to a contraction effect by $K^-$ being introduced into the system. It is noted that such a high density is never seen in nuclear and hypernuclear systems. An essential difference of KNC's from hypernuclei is that an anti-quark, $\bar u$, is implanted into a nuclear system in addition to the $s$ quark through $K^-=(\bar us)$. Thus, the fundamental issues of KNC to be resolved are: 1) "Can an anti-quark, $\bar u$, survive inside a nuclear system of many $u,d$ quarks?" and 2) "Can a boson, $K^-=(\bar us)$, be a constituent of matter?" This paper is a step in an investigation toward solving these issues.

%
\begin{figure}[hbtp]
\begin{center}
\hspace{3.cm}
\vspace{-0.5cm}
\epsfxsize=13cm
\epsfbox{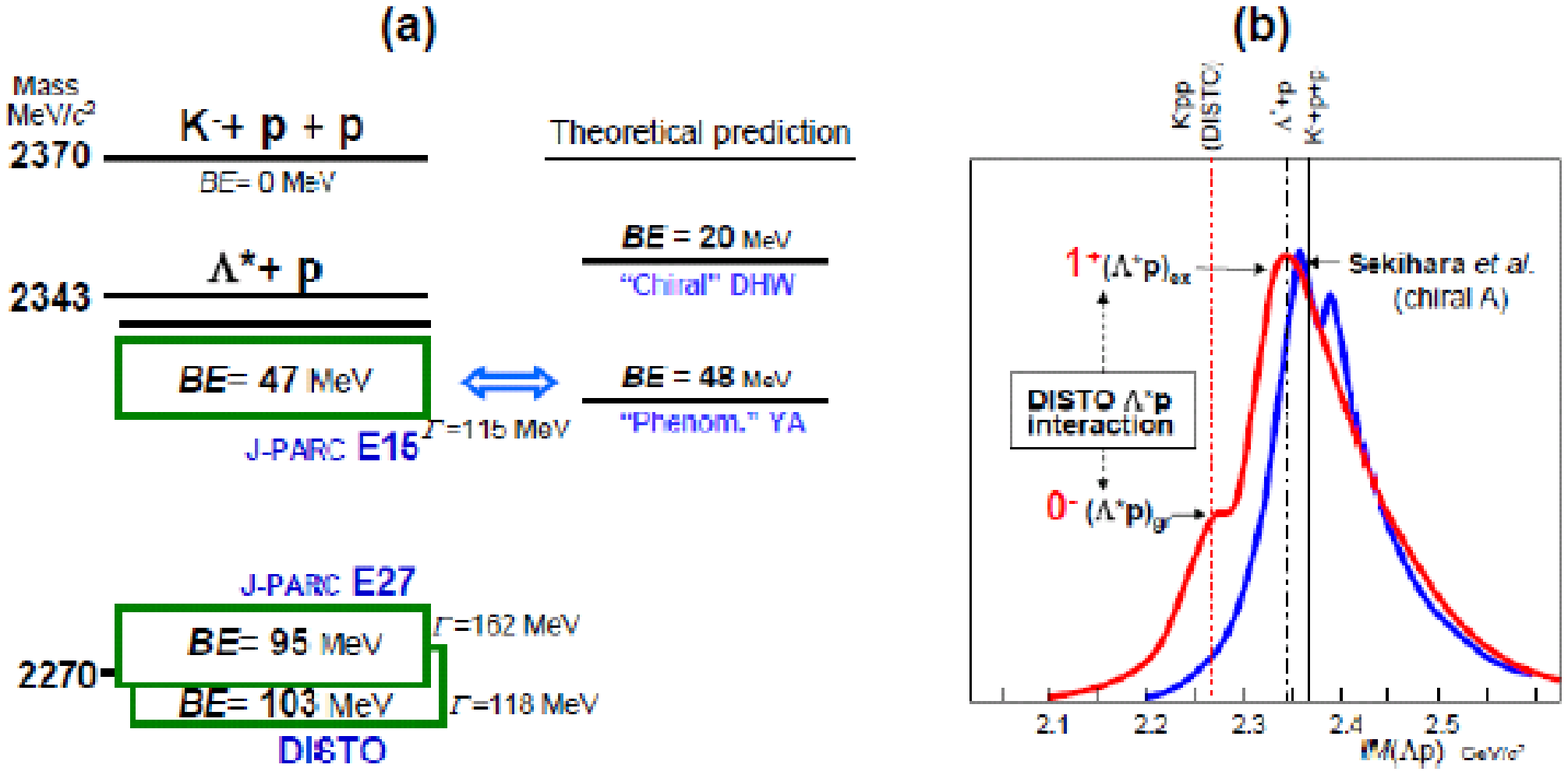}
\vspace{2cm}
\end{center}
\caption{(a) Summary of recent experimental results of $K^-pp$; "DISTO" \cite{Yamazaki10}, "J-PARC E27" \cite{Ichikawa15} and "J-PARC E15" \cite{Ajimura18}. Typical theoretical predictions, "YA" \cite{Yamazaki02} and "DHW" \cite{Dote09}, are also shown. (b) Theoretical $\Lambda p$ invariant-mass spectrum for "J-PARC E15" $^3$He$(K^-,\Lambda p)n$, when "DISTO" $\Lambda^* p$ interaction is used (red curve \cite{Phyo17}). Sekihara {\it et al}.'s "chiral interaction case" \cite{Sekihara16} is shown in blue for comparison. 
\vspace{1cm}}
\label{fig2}
\end{figure}

\section{Experimental evidence for $K^-pp$}

The prediction \cite{Akaishi02,Yamazaki02} of KNC stimulated many theoretical studies \cite{Shevchenko07,Ikeda07,Dote09,Ohnishi17,Dote18} and experimental work \cite{Agnello05,Yamazaki10,Ichikawa15,Ajimura18} concerning the basic $K^-pp$. Theoretical results are roughly grouped into shallow binding ones of $BE \sim 20$ MeV ("chiral $\bar KN$ interaction") and deep binding ones of $BE \sim 50$ MeV ("phenomenological $\bar KN$ interaction"). Figure \ref{fig2}(a) summarizes recent experimental results of $K^-pp$ production with good statistical accuracy accumulated during this decade. They are "DISTO" \cite{Yamazaki10} by $p + p \rightarrow K^+ + X$, "J-PARC E27" \cite{Ichikawa15} by $\pi^+ + d \rightarrow K^+ +X$ and "J-PARC E15" \cite{Ajimura18} by $K^-+^3$He$\rightarrow n+X$, where $X$ includes $K^-pp$ and others with sufficient statistical confidence. 

It is experimentally found in the bound region ($BE > 0$ MeV) that there are two peaks (one at around $BE \sim 50$ MeV) and (the other at around $BE \sim 100$ MeV). Both experimental bindings are deeper than the theoretical "chiral" ones. Although "J-PARC E15" is close to Y-A's "phenomenological" one of Ref. \cite{Yamazaki02}, a serious problem remains as to how well the observed individual peaks correspond to the predicted level energies (the ground and excited states of $K^-pp$). A possible two-peak structure of $K^-pp$ was discussed by W.M.M. Phyo {\it et al}. \cite{Phyo17}, where the DISTO was assigned to be the ground state of the $\Lambda^*p$ system, and thus the $\Lambda^*p$ interaction ("DISTO" interaction) was deduced so. This $K^-pp=(K^-p)^{I=0}p=\Lambda^*p$ system has the $J^\pi=0^-$ ground state at around $BE \sim 100$ MeV and a $J^\pi=1^+$ excited state at around $BE \sim 50$ MeV. The calculated $\Lambda^* p$ invariant-mass spectrum, shown in Fig. \ref{fig2}(b), fits an exclusive datum of $^3$He$(K^-,\Lambda p)n$ from "J-PARC E15" better than the "chiral" of $BE \sim 16$ MeV \cite{Sekihara16}. On the other hand, the statistical significance of the observed $0^-$-state shoulder from the present "E15 data" is weak; high-statistic data is awaited. The small population of the ground state compared with that for the excited state in "J-PARC E15"is intriguing, but can be understood in view of the extremely  unusual large momentum transfer ($Q \sim 1.6$ GeV$/c$) in the DISTO reaction, proposed as "Super Coalescent Reaction" in \cite{Yamazaki07,Hassanvand11,Yamazaki19}.

\vspace{1cm}

\section{Multi-kaonic nuclear clusters}

\subsection{$\Lambda^*$ multiplet}

\begin{figure}[hbtp]
\begin{center}
\hspace{0.5cm}
\vspace{-0.5cm}
\epsfxsize=12cm
\epsfbox{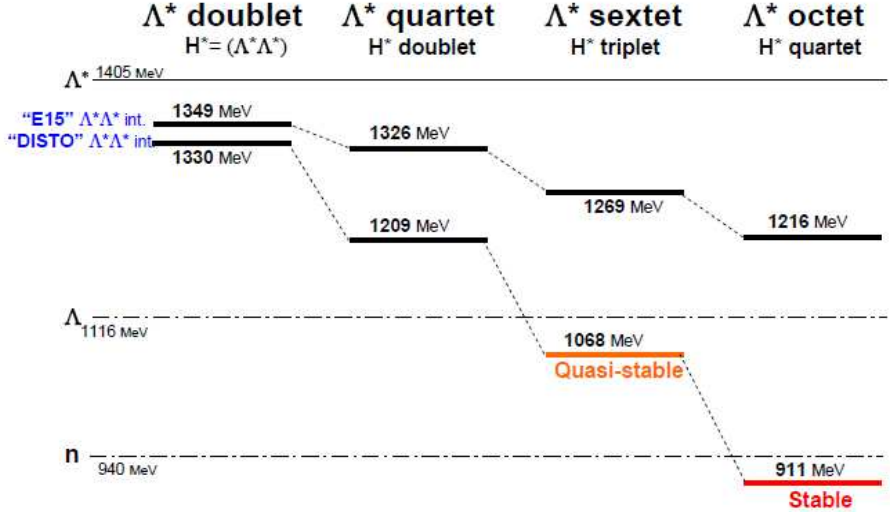}
\vspace{2.cm}
\end{center}
\caption{Effective mass of $\Lambda^*$ in the $(\Lambda^*)_m$ multiplet calculated by the variational method ATMS \cite{Akaishi86} for multiplicity $m= 2, 4, 6, 8$. "DISTO" (or "J-PARC E15") $\Lambda^* \Lambda^*$ interaction is a phenomenological interaction fitted to $BE$ of "DISTO" (or "J-PARC E15").}
\label{fig3}
\end{figure}

Recently, the present authors published a paper on high-density Kaonic-Proton Matter (KPM) composed of $\Lambda^*=(K^-p)$ multiplets \cite{Akaishi17}. Starting from the $\Lambda^*p=(K^-p)-p$ interaction determined so as to reproduce the experimental $BE$ of "DISTO" or of "J-PARC E15", we constructed the $\Lambda^* \Lambda^*=(K^-p)-(K^-p)$ interaction by adding one more kaon to the $(K^-p)-p$, as is explained in Fig. 1 of Ref. \cite{Akaishi17}, based on earlier works of a Heitler-London type formulation by Thida Oo and Myint \cite{Thida10}. The obtained $\Lambda^* \Lambda^*$ attraction bond is almost doubly enhanced compared to the $\Lambda^*p$ one. With this $\Lambda^* \Lambda^*$ interaction, the binding energy, $BE_m$, of the $(\Lambda^*)_m$ multiplet ($m$-body system of $\Lambda^*$) is calculated by the variational method named ATMS \cite{Akaishi86}. The obtained effective mass of $\Lambda^*$ in the $(\Lambda^*)_m$ multiplet, {\it i.e.} $M_{\Lambda^*}^{\rm eff} c^2= M_{\Lambda^*}c^2-BE_m/m$, is given for multiplicity $m= 2, 4, 6, 8$ in Fig. \ref{fig3}. 

As for multi-$\bar K$ nuclei, Gazda {\it et al}. \cite{Gazda08} have conducted Relativistic Mean Field (RMF) calculations, and concluded that the $\bar K$ separation energy (which corresponds to the mass drop of $\Lambda^*$ in the present paper) saturates at a kaon number larger than $\sim 10,$ and does not exceed 200 MeV. This conclusion is unreasonable, because the mean-field model based on any independent-particle motion of the constituents does not fit to the kaonic system, as explained in the following. 

\begin{figure}[hbtp]
\begin{center}
\hspace{+2.cm}
\vspace{-1.5cm}
\epsfxsize=13cm
\epsfbox{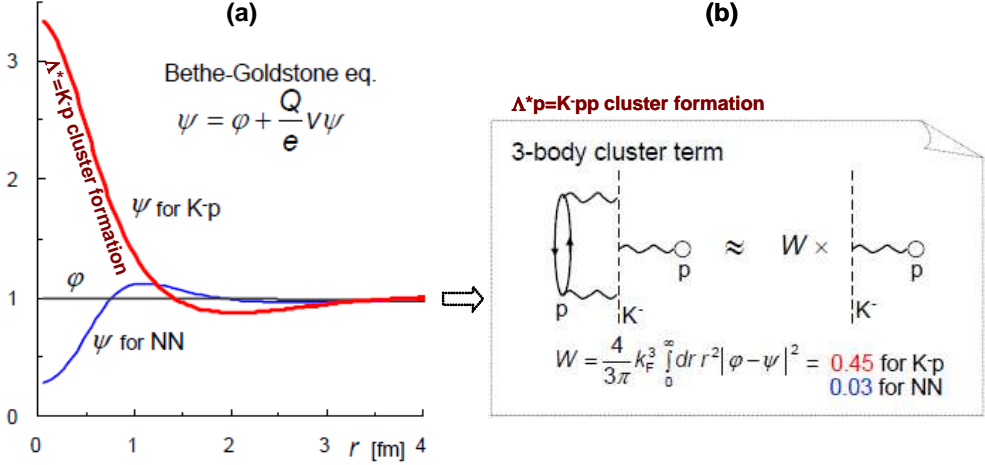}
\vspace{3.cm}
\end{center}
\caption{(a) $\bar KN$ and $NN$ correlated 2-body wave functions in normal-density nuclear medium calculated by using Bethe-Goldstone's equation \cite{Bethe71}. The mean-field model state of $\bar K$ nucleus becomes unstable towards $\Lambda^*=K^-p$ cluster formation. (b)  "Wound" integral, $W$, of the $\bar KN$ wave function and the lowest-order term of $\Lambda^*p=K^-pp$ cluster formation \cite{Bethe71}.}
\label{fig4}
\end{figure}

Figure \ref{fig4}(a) shows wave functions, $\psi$, of $NN$ and $\bar KN$ pairs in nuclear matter at the normal density, which are obtained from the Bethe-Goldstone equation \cite{Bethe71}. It is seen in the case of $NN$ that a "correlation wound"\cite{Gomes58}, which is the difference between the correlated function, $\psi$, and the uncorrelated wave function, $\phi$, is confined within short distances, and heals quickly due to the effect of Pauli exclusion, $Q$, on propagation in nuclear matter. This quick healing assures the independent-particle motion \cite{Gomes58} of $N$ in the nuclear medium. On the other hand, in the case of $\bar KN$ the "wound" is very large and the healing is slow, as seen in Fig. \ref{fig4}(a). The large "wound" is a consequence of very strong $(\bar KN)^{I=0}$ attraction accommodating 27 MeV binding (one-order larger than 2 MeV of $d$), which comes from the additive sum of $\sigma,\omega,\rho$-meson exchange interactions \cite{Muller90}. The large correlation of $\psi$ means that Gazda {\it et al}.'s mean-field model state is unstable towards the 2p-2h (particle-hole excitation) mode of $\Lambda^*=(K^-p)^{I=0}$ cluster formation, though the model state is stable against 1p-1h modes. As denoted in Fig. \ref{fig4}(b), the "wound" integral for $\bar KN$ attains to 0.45, which is 15-times larger than that for $NN$. This means that the mean-field model state is highly unstable also towards the 3p-3h mode \cite{Bethe71} of $\Lambda^*p=K^-pp$ cluster formation. 

As a summary of the above discussion, it is concluded that Gazda {\it et al}.'s mean-field model state of multi-$\bar K$ nuclei is unstable towards the formation of $\Lambda^*$ and the successive formation of $(\Lambda^*)_m$ multiplets given in Fig. \ref{fig3}. Then, the most striking result of Fig. \ref{fig3} is that the $\Lambda^*$ in $(\Lambda^*)_m$ becomes more stable than the $\Lambda$ particle at $m=6$, and than the lightest-mass baryon of $(n)_8$ at $m=8$, in the case of the "DISTO" interaction. 

\vspace{1cm}

\subsection{Kaonic proton matter (KPM) composed of $\Lambda^*=(K^-p)$ multiplets}

As an instructive example, we discuss unique properties of the stable $(\Lambda^*)_8$ multiplet of the "DISTO" interaction case. The total mass of the $\Lambda^*$ multiplet, $M((\Lambda^*)_m)c^2= M_{\Lambda^*}c^2~m-BE_m$, shown in Fig. \ref{fig3}, can be expressed approximately as $M((\Lambda^*)_m)c^2 \sim 1405 m -135 m(m-1)/2$ in MeV for $m=2$ to 8. Then, the binding energy per $\Lambda^*$ is $BE_m/m \sim 135 (m-1)/2 \sim 473$ MeV for $m=8$, and the $\Lambda^*$ separation energy from $(\Lambda^*)_m$ is $S_m(\Lambda^*) \sim 135 (m-1) \sim 945$ MeV for $m=8$.

The $(\Lambda^*)_8$ is very tight and dense with a binding per $\Lambda^*$ of $\sim 473$ MeV, compared to ordinary nuclei with a binding per $N$ of $\sim 8$ MeV. It is noted that the $S_m(\Lambda^*)$ is 2-times larger than the $BE_m/m$ due to a  rearrangement of $(\Lambda^*)_{m-1}$ after the separation of a $\Lambda^*$. Because of this large rearrangement energy, the $(\Lambda^*)_m$ system increases its stiffness. The separation energy, $S_8(\Lambda^*)$, attains to almost $\sim 1$ GeV, which is 2-orders of magnitude larger than $\sim 8$ MeV of $N$ separation from ordinary nuclei. Thus, the $(\Lambda^*)_8$ is extremely stiff, and any decay of $\Lambda^*$, such as $\Lambda^* \rightarrow p+e^-+{\bar \nu}$, is energetically prohibited inside the system. These features are completely different from those of ordinary nuclei, and suggest a new paradigm of particle and nuclear physics with $\Lambda^*=(K^-p)$. The $(\Lambda^*=(K^-p))_8$ multiplet is characterised by unique properties of the "neutral-dense-stiff-stable" nature, where an anti-quark $\bar u$ survives through $\Lambda^*=(K^-p)= ((\bar us)(uud))$, as revealed by a recent lattice QCD calculation \cite{Hall15}.
%
\begin{figure}[hbtp]
\begin{center}
\hspace{0.5cm}
\vspace{-0.5cm}
\epsfxsize=9cm
\epsfbox{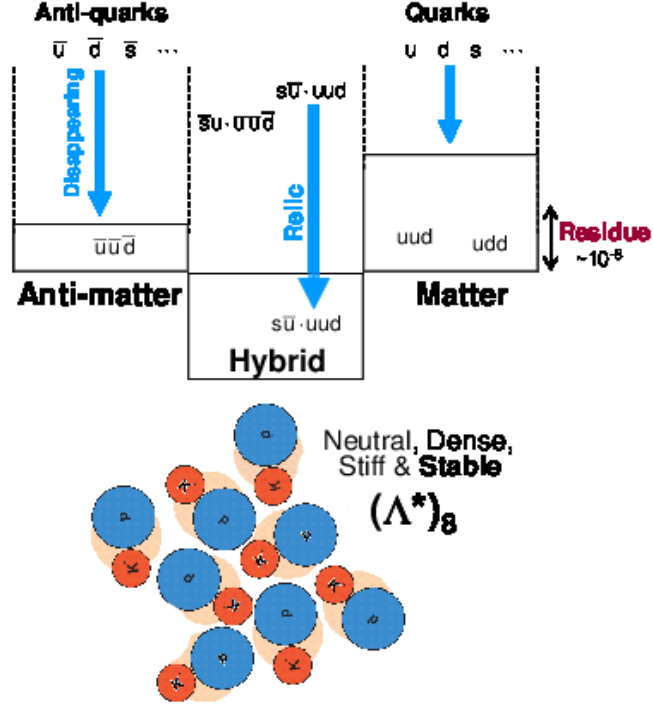}
\vspace{2.0cm}
\end{center}
\caption{Schematic picture of a possible scenario, suggested from the $(\Lambda^*)_8$ multiplet, of $\bar qq$-hybrid bound-state formation at early stage of the universe.}
\label{fig5}
\end{figure}

Figure \ref{fig5} is a schematic picture of a possible scenario of $\bar qq$-hybrid bound-state formation at an early stage of the universe, as suggested from the stable multiplet of $(\Lambda^*)_8=((\bar us)(uud))_8$. At the baryogenesis stage of the early universe a substantial amount of primordial anti-quarks were captured into kaonic proton matter (KPM) composed of $\Lambda^*=(\bar us)(uud)$ multiplets, where the anti-quarks found places to survive forever. KPM might be a relic of disappearing anti-quarks, playing an essential role as hidden components in the universe.

Finally, two comments are added on the stable $(\Lambda^*)_8$.  1) In a dilute gas of $(\Lambda^*)_8$'s two-body collisions occur in repulsive way, since too-strong attractive interaction is equivalent to repulsive interaction in scattering states. 2) In neutron-star matter the transition of an $(n)_8$ aggregate to $(\Lambda^*)_8$ is extremely suppressed, because the $n \rightarrow \Lambda^*$ weak transition must take place 8 times within the $\Lambda^*$ lifetime of $10^{-23}$ sec. 

\section{Is $\Lambda^*$ matter stable or unstable?}

\begin{figure}[hbtp]
\begin{center}
\hspace{1.5cm}
\vspace{-1.5cm}
\epsfxsize=13.cm
\epsfbox{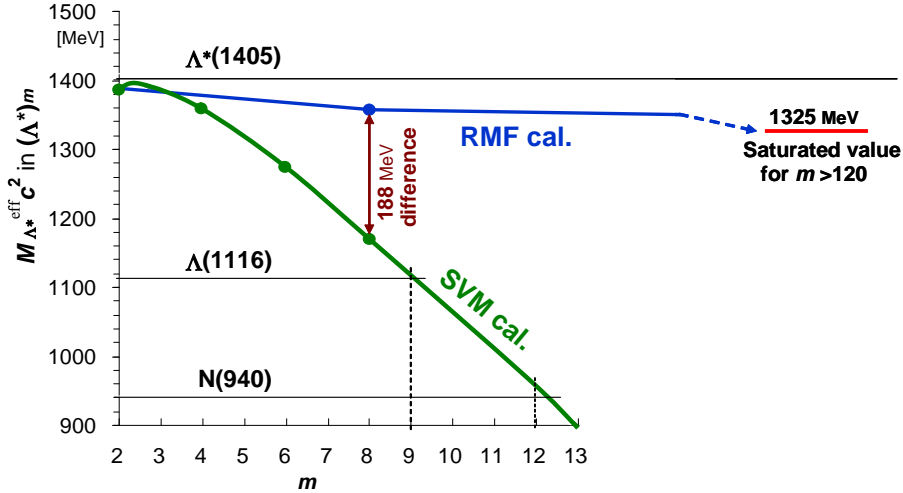}
\vspace{2.5cm}
\end{center}
\caption{Stochastic Variational Method (SVM) calculation by Nemura \cite{Nemura18} for the effective $\Lambda^*$ mass in $(\Lambda^*)_m$, $M_{\Lambda^*}^{\rm eff} c^2= M_{\Lambda^*}c^2-BE_m/m$. Nemura's mass-formula curve well reproduces the SVM results. The Relativistic Mean Field (RMF) calculation 
of $(\Lambda^*)_{m=8 \sim 168}$ by Hrt$\acute{\rm a}$nkov$\acute{\rm a}$ {\it et al}. \cite{Hrtankova18} is indicated for comparison.}
\label{fig6}
\end{figure}

As for  $\Lambda^*$ matter, Hrt$\acute{\rm a}$nkov$\acute{\rm a}$ {\it et al}. \cite{Hrtankova18} calculated the $BE_m/m$ of $(\Lambda^*)_m$ system from $m = 8$ to 168 by a $\sigma$-$\omega$ Relativistic Mean Field (RMF) approach \cite{Horowitz81}. The curve, referred to as "RMF cal." in Fig. \ref{fig6}, indicates their main result obtained in the case of the Machleidt potential \cite{Machleidt89} with the $\omega \Lambda^*$ coupling adjusted so as to get $BE_2= 40$ MeV for $(\Lambda^*)_2$. They found that   $BE_m/m$ saturates for $m > 120$ with a value of $\sim 80$ MeV ("RMF limit"), which is far less than $\sim 473$ MeV of the $(\Lambda^*)_8$ discussed in Subsec. 3.2, they highly insisted that "the $\Lambda^*$ matter is totally unstable, and the $\Lambda^*$ stable-matter scenario promoted by Akaishi-Yamazaki (AY's scenario) \cite{Akaishi17} is unlikely to be substantiated in standard many-body schemes".  
%
%
\begin{figure}[hbtp]
\begin{center}
\hspace{-1.5cm}
\vspace{0.5cm}
\epsfxsize=10cm
\epsfbox{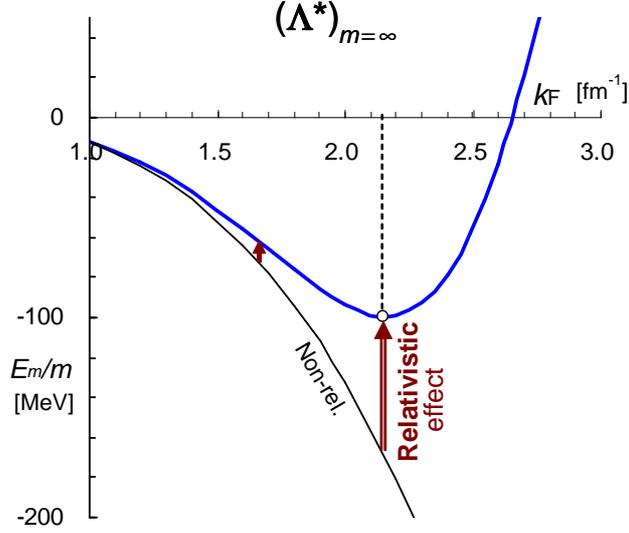}
\vspace{1cm}
\caption{Relativistic Mean Field calculation for $m=\infty$. The relativistic effect depends largely on the baryon density ($\rho_{\rm B} \propto k_{\rm F}^3)$. The effect is estimated to be on the order of 10 MeV for $(\Lambda^*)_8$ within a local-density approximation.}
\label{fig7}
\end{center}
\end{figure}

On the other hand, Nemura \cite{Nemura18} has treated  $m$-body $\Lambda^*$ systems for $m = 4 \sim 8$ with the same adjusted Machleidt potential by the Stochastic Variational Method (SVM), while taking account of full anti-symmetrization and higher-order correlations. The results are shown as "SVM cal." in Fig. \ref{fig6}. There is a substantial discrepancy between the curve of "SVM cal." and that of "RMF cal.". At $m=8$, the SVM value attains to $BE_m/m=238$ MeV, far exceeding the "RMF limit" of $\sim 80$ MeV, though the RMF value stays 50 MeV. The final value of RMF at $m=\infty$ can be easily calculated by using the original RMF of Walecka \cite{Walecka74}. The value is 87 MeV at $k_{\rm F}=2.15$ fm$^{-1}$, as is given in Fig. \ref{fig7}, and its breakdown consists of $\langle T \rangle = 38$ MeV, $\langle V_{\omega} \rangle _{\rm Born}\rho_{\rm B}= 316$ MeV and $\langle V_{\sigma} \rangle _{\rm Born}\rho_{\rm s}= -441$ MeV. It should be noticed that the interaction energy of Hrt$\acute{\rm a}$nkov$\acute{\rm a}$ {\it et al}.'s RMF calculation is counted also in a Born (and Hartree) approximation, by disregarding higher-order contributions of the strongly attractive $\Lambda^* \Lambda^*$ potential adjusted to $BE_2=40$ MeV. This $\langle V \rangle _{\rm Born}\rho$ treatment is a serious shortcoming of the RMF calculation applied to   $\Lambda^*$ matter. 

The discrepancy between "RMF" and "SVM" is already $238-50=188$ MeV at $m=8$. In order to know how large is the relativistic effect, a non-relativistic curve calculated by replacing $\rho_{\rm s} \rightarrow \rho_{\rm B}$ is shown in Fig. \ref{fig7}. The relativistic effect is 63 MeV at $m=\infty$, and is estimated to be roughly 10 MeV at $m=8$ in a local-density approximation. Thus, the majority of the 188 MeV difference comes not from the relativistic effect, but from multi-cluster correlations among $\Lambda^*$'s. A lack of the multi-cluster contributions is a fatal defect of Hrt$\acute{\rm a}$nkov$\acute{\rm a}$ {\it et al}.'s RMF calculation. Nemura's SVM calculation fully includes these multiple correlations.

Nemura's mass-formula curve fitted to the accurate SVM results at $m = 4 \sim 8$ predicts that the $(\Lambda^*)_m$ multiplet becomes absolutely stable at around $m=12$, even when the adjusted Machleidt potential is employed. Thus, the SVM calculation with full higher-order correlations clearly denies Hrt$\acute{\rm a}$nkov$\acute{\rm a}$ {\it et al}.'s negative claims, but certainly does support AY's $\Lambda^*$ stable-matter scenario.

\section{Concluding remarks}

 $\Lambda^* = \Lambda (1405)=(K^-p)^{I=0}$ plays an essential role in forming kaonic nuclear clusters (KNC).  Production data of the $K^-pp=\Lambda^*p$ have been accumulated by DISTO, J-PARC E27 and J-PARC E15 experiments. Two peak structures are observed at around $BE \sim 50$ MeV and $BE \sim 100$ MeV. Although a problem remains as to how to reconcile the two peaks to specific configurations of $K^-pp$, further experimental searches will produce clear indications, concerning  super-strong $\Lambda^*p$ attraction. In the case of $K^-K^-pp$ the $\Lambda^* \Lambda^*$ covalent bond due to $K^-$ migration is doubly enhanced compared to the $\Lambda^*p$ one. As a consequence, $m$-body $\Lambda^*$ multiplet, $(\Lambda^*)_m$, with $m(m-1)/2$ attractive covalent bonds becomes more stable than neutron aggregate, $(n)_m$, at $m=8 \sim 12$. Hrt$\acute{\rm a}$nkov$\acute{\rm a}$ {\it et al}.'s objection \cite{Hrtankova18} against stable $\Lambda^*$ matter is clearly denied by Nemura's SVM calculation \cite{Nemura18}. The present theoretical analysis suggests the possible existence of stable $\Lambda^*=(K^-p) = ((\bar us)(uud))$ matter in the quark-antiquark hybrid sector. 
%

\begin{figure}[hbtp]
\begin{center}
\hspace{0.5cm}
\vspace{-0.5cm}
\epsfxsize=13cm
\epsfbox{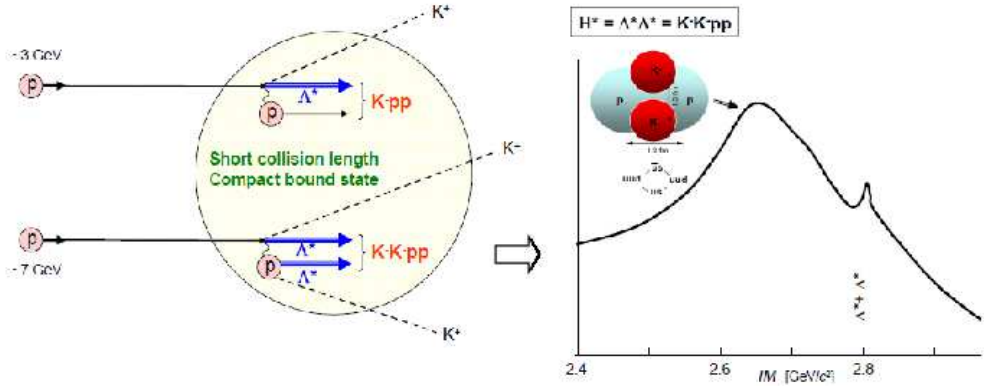}
\vspace{2.5cm}
\end{center}
\caption{Possible formation of $K^-K^-pp=\Lambda^* \Lambda^*$ from high-energy $pp$ collision and its mass spectrum calculated by Hassanvand {\it et al}. \cite{Hassanvand11}.}
\label{fig8}
\end{figure}

The production of $K^-K^-pp=\Lambda^* \Lambda^*$ \cite{Hassanvand11} by means of high-energy $pp$ or heavy-ion collisions is awaited as a doorway to a new paradigm of particle and nuclear physics concerning so-far unknown $\Lambda^*$ matter \cite{Yamazaki19}.

\section*{Acknowledgments}

The authors would like to thank Prof. Khin Swe Myint and members of Mandalay Nuclear Theory Group for fruitful collaborations and Dr. H. Nemura for the elaborate SVM calculations of the $(\Lambda^*)_m=4,6,8$ multiplets given in Sec. 4.

\end{document}